\documentclass[preprints,review,accept,oneauthor,pdftex]{Definitions/mdpi} 

\firstpage{1} 
\pubvolume{xx}
\issuenum{1}
\articlenumber{5}
\pubyear{2020}
\copyrightyear{2020}
\history{Received: date; Accepted: date; Published: date}

\Title{Neutron-Capture Element Abundances in Planetary Nebulae}

\Author{N.\ C.\ Sterling $^{1}$}

\AuthorNames{N .\ C.\ Sterling}

\address{%
$^{1}$ \quad University of West Georgia, Carrollton, GA 30118, USA, nsterlin@westga.edu}

\corres{Correspondence: nsterlin@westga.edu}

\abstract{Nebular spectroscopy is a valuable tool for assessing the production of heavy elements by slow neutron(\emph{n})-capture nucleosynthesis (the \emph{s}-process).  Several transitions of \emph{n}-capture elements have been identified in planetary nebulae (PNe) in the last few years, with the aid of sensitive high-resolution near-infrared spectrometers.  Combined with optical spectroscopy, the newly discovered near-infrared lines enable more accurate abundance determinations than previously possible, and provide access to elements that had not previously been studied in PNe or their progenitors.  Neutron-capture elements have also been detected in PNe in the Sagittarius Dwarf galaxy and in the Magellanic Clouds.  In this brief review, I discuss developments in observational studies of \emph{s}-process enrichments in PNe, with an emphasis on the last five years, and note some open questions and preliminary trends.}

\keyword{planetary nebulae: general; nuclear reactions, nucleosynthesis, abundances; stars: AGB and post-AGB; infrared: general}

\begin{document}


\section{Introduction}
Neutron(\emph{n})-capture elements (atomic number $Z>30$) are produced by slow and rapid \emph{n}-capture nucleosynthesis (the \emph{s}- and \emph{r}-processes, respectively).  These processes are distinguished by the relative time scales of successive \emph{n}-captures and $\beta$-decays of the participating isotopes, and produce distinct enrichment patterns, with the \emph{r}-process leading to more neutron-rich nuclei \cite{sneden08}.  Despite their cosmic rarity, \emph{n}-capture element abundances can reveal valuable details of the chemical evolution and star formation histories of galaxies \cite[e.g.,][]{frebel_norris15, skuladottir20}, and therefore it is critical to understand their sites of origin and chemical yields in stars of different mass and metallicity.

The \emph{r}-process is known to occur in high-mass stars \cite[e.g.,][]{sneden08}, and the discovery of lanthanide-rich material in the kilonova AT~2017gfo \cite{tanvir17} established that neutron star mergers produce \emph{r}-process nuclei.  However, only bulk lanthanide abundances can presently be determined in kilonovae, with the possible exception of Sr \cite{watson19}, and a lack of atomic data exacerbates this issue \cite{smartt17}.  It is also unclear whether the \emph{r}-process occurs in other astrophysical sites \cite[e.g.,][]{cote19}.  This underlines the need to better understand the production of elements by the \emph{s}-process, which contributes to the abundances of the same elements (though often different isotopes) formed by the \emph{r}-process.

The \emph{s}-process takes place in low- and intermediate-mass stars (1--8~M$_{\odot}$) during the thermally-pulsing asymptotic giant branch (AGB) phase of evolution.  Free neutrons are produced by $\alpha$-captures onto $^{13}$C in the intershell region between the H- and He-burning shells.  Iron-peak nuclei undergo a series of \emph{n}-captures interlaced with $\beta$-decays to form heavier elements, which are transported to the surface (along with C from partial He burning) by convective dredge-up events.  A second source of neutrons, $\alpha$-captures onto $^{22}$Ne, requires higher temperatures and does not substantially affect \emph{s}-process enrichment patterns except in more massive ($>4-5$~M$_{\odot}$) stars \cite{busso99, karakas_lattanzio14}.  These more massive stars produce Type~I PNe \cite{peimbert78}, which exhibit N and He enrichments due to the operation of the CNO cycle at the base of the convective envelope (``hot bottom burning,'' or HBB) \cite{karakas_lattanzio14}. 

P\'equignot \& Baluteau \cite{pb94} first identified \emph{n}-capture element emission lines in an ionized nebula, the bright PN NGC~7027.  Although their spectrum did not resolve some transitions from weak features of more abundant elements, Sharpee et al.\ \cite{sharpee07} later confirmed the identity of lines from elements including Br, Kr, Rb, Xe, and possibly Ba.  The work of P\'equignot \& Baluteau inspired Dinerstein \cite{dinerstein01} to realize that two long-unidentified near-infrared (NIR) features at 2.1980 and 2.2864~$\mu$m\footnote{Vacuum wavelengths are used in this paper for NIR lines and air wavelengths for optical transitions.} are fine-structure transitions of [Kr~III] and [Se~IV], respectively.  These Kr and Se lines have since been detected in more than 100 PNe in the Galaxy and nearby galaxies \cite{sterling_dinerstein08, mashburn16}.

Nebular spectroscopy uniquely probes aspects of heavy element production.  PNe form after the cessation of nucleosynthesis, whereas AGB stars abundances may be altered by additional dredge-up events through the end of the AGB phase.  The compositions of PNe can thus be used to determine chemical yields, which are sensitive to poorly-understood processes in the progenitor AGB stars, including mass loss, convective mixing, convection, and rotation \cite{karakas09, karakas_lattanzio14, sterling16}.  In addition, many of the elements detected in AGB stars (e.g., Sr, Y, Zr, La, and Ba) are highly refractory \cite{lodders03} and are likely depleted into dust grains in PNe.  Nebular spectroscopy provides access to the noble gases Kr and Xe, as well as other elements (e.g., Se, Br, Cd, Te) that are not detectable in AGB stars.  Finally, the systematic uncertainties in nebular abundances, which include corrections for unobserved ionization states and depletion into dust, are independent of those involved in stellar spectroscopy.

In this review, I discuss observational studies of \emph{s}-process enrichments in PNe, with emphasis on the past five years, and note some trends and open questions.  Newly-identified emission lines have spurred investigations into the atomic data needed to interpret these features, work that has developed concurrently with observations \cite[see][for a review]{sterling17a}.

\section{Near-Infrared Observations}\label{nir}

The NIR spectral region hosts transitions of many heavy element ions that either cannot be detected or are severely compromised by blends at optical wavelengths.  Combined with optical data, NIR spectroscopy enables more accurate \emph{n}-capture element abundance determinations, which are critical for constraining AGB and chemical evolution models.

Several NIR emission lines from \emph{n}-capture elements, including Ge ($Z=32$), Se (34), Br (35), Kr (36), Rb (37), Cd (48), and Te (52), have been identified recently in PNe.  These discoveries were made possible by the advent of sensitive high-resolution NIR spectrometers.  For each feature discussed below, atomic data (i.e., transition probabilities and effective collision strengths) have been calculated.

Five emission lines have been discovered with the Immersion Grating INfrared Spectrometer (IGRINS) \cite{park14} on the 2.7-m Harlan J.\ Smith Telescope at McDonald Observatory.  IGRINS provides simultaneous coverage of the \textit{H} and \textit{K}~bands (1.45--2.45~$\mu$m) at a spectral resolution $R=45,000$.  Sterling et al.\ \cite{sterling16} used IGRINS to identify [Rb~IV]~1.5973~$\mu$m, [Cd~IV]~1.7204~$\mu$m, and [Ge~VI]~2.1930~$\mu$m in the PNe NGC~7027 and IC~5117.  The rest wavelengths of the [Rb~IV] and [Cd~IV] transitions are near sky emission features, which can mask the nebular lines in low dispersion spectra.  Madonna et al.\ \cite{madonna18} later made use of IGRINS to identify [Br~V]~1.6429 and [Te~III]~2.1019~$\mu$m in IC~418 and NGC~7027.

These investigations highlight two important points in identifying \emph{n}-capture element transitions.  First, high resolution is often critical, both to resolve the lines from nearby features, and to assess other possible identifications.  If the line has an alternate identification, other transitions from the same multiplet and/or arising from the same upper level should be detectable, but these may have close wavelength coincidences with other features.  Secondly, it is crucial to utilize recent energy level determinations.  As an example, the energy levels from the compilation of Moore (1958) \cite{moore58} give a wavelength of 2.1048~$\mu$m for [Te~III], nearly 30~\AA\ from more recent values \cite{tauheed_naz11}!  This underlines the key role that laboratory spectroscopy plays in the study of \emph{s}-process enrichments in PNe.

Some \emph{n}-capture element transitions can be identified using lower resolution spectra, as in the case of [Se~III] 1.0994 and [Kr~VI]~1.2330~$\mu$m \cite{sterling17b}.  These lines were detected with the Folded-Port InfraRed Echellette (FIRE) spectrometer \cite{simcoe13} on the 6.5-m Baade Telescope at Las Campanas Observatory.  [Se~III] can be resolved from nearby blends at the FIRE resolution of 5000--8000, but caution must be used in identifying the 1.2330~$\mu$m feature, as [Fe~VI], H$_2$~3-1~S(1), and N~I lines share the same wavelength.

Additional high-resolution instruments have become available recently, including the iSHELL echelle spectrometer \cite{rayner16} on the 3-m NASA Infrared Telescope Facility (IRTF), and the Habitable Planet Finder (HPF) \cite{mahadevan14} on the 10-m Hobby Eberly Telescope.  The HPF ($R=55,000$) provides spectral coverage for wavelengths 0.82--1.28~$\mu$m, while iSHELL covers 1.1--5.3~$\mu$m at resolutions up to 75,000.

\begin{figure}[H]
\centering
\includegraphics[width=12 cm]{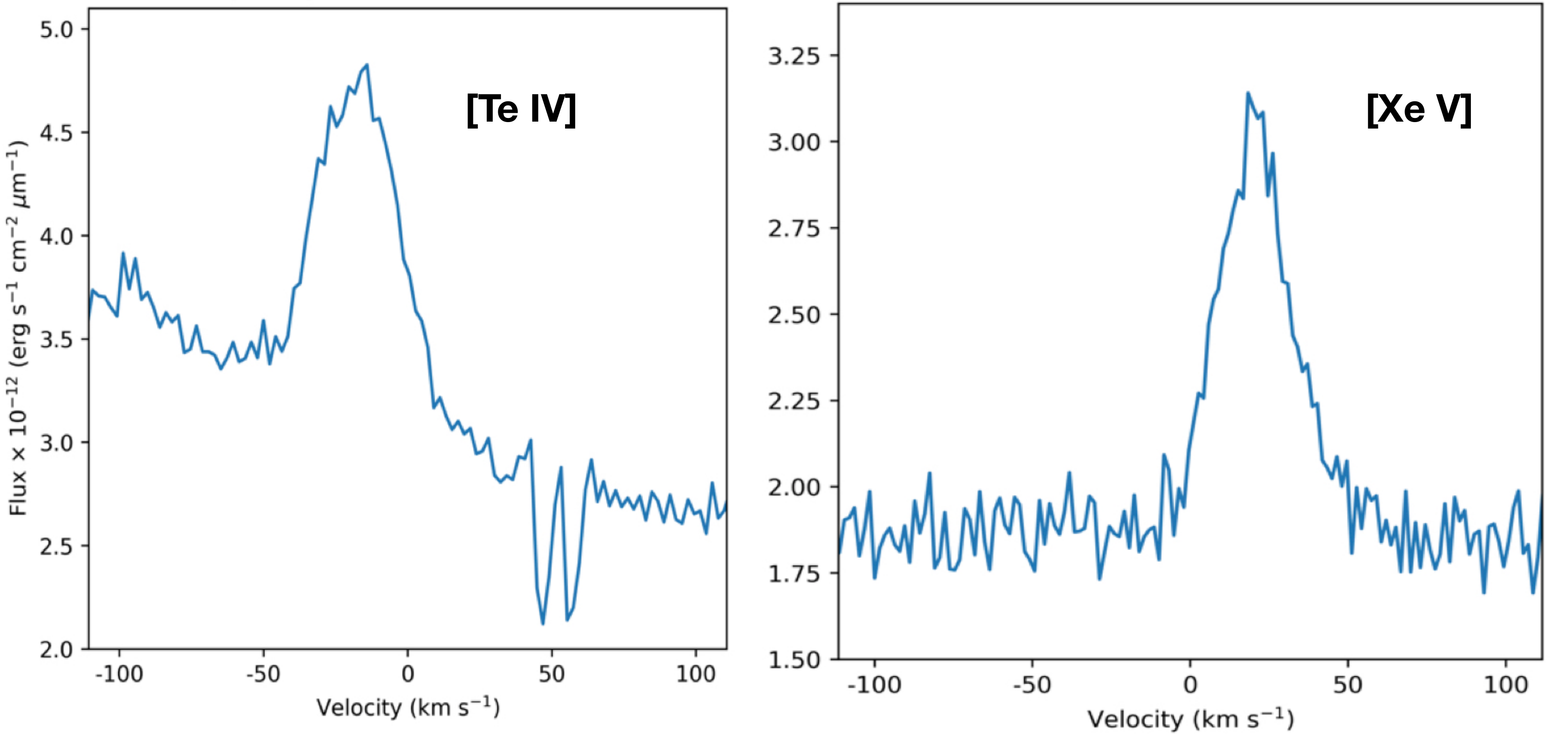}
\caption{Detections of [Te~IV]~1.0843 and [Xe~V]~1.0762~$\mu$m in IC 5117, based on preliminary results from the HPF spectrometer on the 10.4-m Hobby-Eberly Telescope (Dinerstein et al.\ 2020b, in preparation).  The absorption features near [Te~IV] are due to imperfect sky subtraction.}\label{figure}
\end{figure}

Using iSHELL, Dinerstein et al.\ (2020a, in preparation) identified [Rb~III]~1.3560~$\mu$m, the only collisionally-excited transition of Rb$^{2+}$, in two PNe.  This line lies just past the $J$~band atmospheric window, and can only be detected on low water vapor nights with the IRTF (elevation 4200~m) or from space.  Figure~\ref{figure} shows [Te~IV] and [Xe~V] features detected with the HPF in IC~5117 (Dinerstein et al. (2020b, in preparation), the first detection of each of these ions.  The [Te~IV] transition, on the red wing of the strong He~I~1.0833~$\mu$m line, is of particular import since the only other Te ion detected, Te$^{2+}$, is likely to have a low ionic fraction based on its ionization potential of 27.8~eV.

The number of newly-detected lines demonstrates that NIR spectroscopy is particularly valuable for assessing the impact of \emph{s}-process enrichments in PNe.  Observing ionization stages that cannot be detected in the optical enables more accurate abundance determinations.  In the optical, only [Rb~IV], [Rb~V], and [Rb~VI] are detectable \cite{pb94, sharpee07}.  The latter two are trace species due to their high ionization potentials, and [Rb~IV]~5759.55~\AA\ can be blended with a weak He~II Pfund line.  The isolated [Rb~III] and [Rb~IV] NIR lines sample two of the dominant Rb ions in most PNe.  Since Rb enrichments are sensitive to the \emph{s}-process neutron density, the Rb abundance relative to other \emph{n}-capture elements can be used to diagnose the neutron source in the progenitor star (see Sec.\ \ref{openqs}) \cite{garcia-hernandez09, vanraai12, karakas_lugaro16}.

Cd and Te join Xe \cite{pb94} as the only non-refractory elements beyond the first \emph{s}-process peak detected in PNe.  This peak, at $Z=38$--40 (Sr, Y, and Zr), is due to isotopes with a closed neutron shell ($N=50$).  Such neutron-magic nuclei have low \emph{n}-capture cross sections, and the relative abundances of elements on either side of the peak is sensitive to the time-averaged \emph{s}-process neutron flux \cite{busso99}.  Cd and Te thus provide important new constraints to theoretical predictions. Indeed, a sufficient number of \emph{n}-capture elements have been detected in individual PNe for meaningful comparisons to model predictions \cite[e.g.,][]{sterling16}.

\section{Optical Spectroscopy}\label{opt}

Several \emph{n}-capture elements have optical transitions, including species that are not detectable in the NIR.  Deep, high-resolution ($R>20,000$) spectra are needed to unambiguously identify these lines.

Very deep spectra of PNe have been acquired with the UVES spectrograph \cite{dodorico00} ($R=45,000$) on the 8.2-m Very Large Telescope (VLT). Garc\'ia-Rojas et al.\ observed NGC~3918 \cite{garcia-rojas15}, and detected more than 750 emission lines, including [Kr III--V], [Xe~III--IV], and [Rb~III--IV].  When combined with the [Kr~VI]~1.2330~$\mu$m line detected by \cite{sterling17b}, all significantly populated Kr ions have been detected in NGC~3918, with the exception of Kr$^+$, whose sole collisionally-excited transition at 1.8622~$\mu$m is accessible only by space observatories.  The full set of Kr ionization correction prescriptions of \cite{sterling15} were used to derive the Kr abundance ([Kr/O]~=~0.66$\pm$0.09).  The Se abundance was derived from [Se~III]~$\lambda$8854.00, but blending with He~I emission at 8854.20~\AA\ rendered it uncertain.  The resulting [Kr/Se]~=~0.4$\pm$0.2 in NGC~3918 agrees with the typical values of 0.5$\pm$0.2~dex from NIR spectra \cite{sterling15}, but is larger than the value of [Kr/Se]~=~0.1--0.2 dex predicted by models \cite[e.g.,][]{karakas_lugaro16}.  Given the accuracy of Kr abundances, this discrepancy with theoretical predictions may be due to inaccurate Se abundances.  

Madonna et al.\ \cite{madonna17} acquired a deep UVES spectrum of the moderate-excitation PN NGC~5315, detecting [Se~III], [Br~III], [Kr~III--IV], and possibly [Xe~IV].  While this PN is not \emph{s}-process enriched, two points regarding \emph{n}-capture element abundances can be taken from this study.  The Se abundance was derived from the unblended [Se~III]~1.0994~$\mu$m line \cite{sterling17b} and [Se~IV]~2.2864~$\mu$m, allowing for the first empirical test of Se ionization correction schema \cite{sterling15}.  When derived from only [Se~IV], the Se abundance is $\sim$0.5~dex smaller than when Se$^{2+}$ is incorporated into the abundance derivation \cite{sterling17b}.  This suggests that Se abundances computed from [Se~IV] may be underestimated, providing a possible remedy to the discrepancy between empirical and theoretical [Kr/Se] ratios.  Additional observations of [Se~III]~1.0994~$\mu$m are needed to test the accuracy of existing Se abundances derived from [Se~IV].  Secondly, Madonna et al.\ found that the Br abundance derived from [Br~III] 6556.56~\AA\ is unphysically large, when compared to the upper limit derived from [Br~III] $\lambda$6130.40, suggesting that this line is either contaminated by unidentified features or is incorrectly identified.  Furthermore, the $\lambda$6130.40 line can be contaminated by C~III \cite{sharpee07, garcia-rojas15}, and optical [Br~IV] lines have been detected only in NGC~7027 \cite{sharpee07}.  Therefore Br abundances from optical spectroscopy should be regarded with caution at this time.

Otsuka \& Hyung \cite{otsuka_hyung20} observed the high-excitation, low-metallicity PN J900 with the Bohyunsan Echelle Spectrograph ($R=43,000$) \cite{kim02} on the 1.8-m telescope at Bohyunsan Optical Astronomy Observatory.  They found that Se, Kr, and Rb are enhanced relative to Ar in J900 by approximately an order of magnitude, while [Xe/Ar] is 50 times higher than solar.  Fluorine, which can also be produced during the AGB \cite[e.g.,][]{karakas_lugaro16}, was also found to be enriched.  The F and \emph{n}-capture element abundances agree well with model predictions for $\sim$2.0--2.5~M$_{\odot}$ progenitors \cite{karakas18}.

Aleman et al.\ \cite{aleman19} used XSHOOTER \cite{vernet11} on the VLT to observe the fullerene-rich PN Tc~1.  They derived a large Kr relative to solar (0.85~dex) from [Kr~III]~6826.70~\AA\ (see Sec.\ \ref{openqs}).

\section{Extragalactic Planetary Nebulae}\label{exgal}

Recent observations using large ($>6$-m) telescopes demonstrate that \emph{n}-capture elements can be detected in PNe belonging to nearby galaxies.  Such investigations enable \emph{s}-process enrichments to be studied in a wider range of metallicities and progenitor masses than are found in Galactic populations.  

Otsuka and collaborators first published studies of \emph{n}-capture elements in extragalactic PNe, with optical Subaru High Dispersion Spectrograph \cite{noguchi02} observations of three objects in the Sagittarius dwarf galaxy.  In Hen~2-436 and Wray~16-423, Kr is enhanced relative to solar by a factor of $\sim$7 \cite{otsuka11, otsuka15}.\footnote{Enhanced abundances of Se and Kr in Hen 2-436 indicated by their NIR lines were first reported by Wood et al.\ \cite{wood06}.}  In BoBn~1, [Kr~IV], [Xe~III], and possibly [Rb~V] were detected, but only limits could be placed on the ionic and elemental abundances due to blending ([Xe~III]~$\lambda$5846.77 with He~II) or low signal-to-noise.  The upper limits allow for enrichments of Kr and Xe, as may be indicated by the large F abundance.

Mashburn et al.\ \cite{mashburn16} conducted the first study of \emph{n}-capture element abundances in Magellanic Cloud PNe, observing seven LMC and three SMC PNe with FIRE on the 6.5-m Baade Telescope and the Gemini Near-InfraRed Spectrograph (GNIRS) on the 8.1-m Gemini-South telescope.  They detected [Kr~III]~2.1980 and [Se~IV]~2.2864~$\mu$m in eight of the 10 objects.  Kr was found to be enriched by 0.6--1.3~dex in the six PNe in which it was detected, while Se was found to be enriched by 0.5--0.9~dex in five of the seven objects exhibiting [Se~IV] emission.

The fact that light \emph{n}-capture elements such as Se and Kr are substantially supersolar in Sagittarius Dwarf and Magellanic Cloud PNe suggests that heavier elements may be enriched by even larger amounts.  AGB nucleosynthesis models predict that the \emph{s}-process at low metallicity will result in higher abundances of elements near the second \emph{s}-process peak ($N=82$, at Ba, La, and Ce) relative to the first peak, due to the paucity of iron-peak ``seed'' nuclei \cite{busso99, karakas_lattanzio14}.  Deeper observations are needed to detect elements beyond the first \emph{s}-process peak in extragalactic PNe.

\section{Discussion}\label{openqs}

In this section, I discuss trends and open questions related to the observations described above.

\textbf{Fullerene PNe.}  Infrared emission features from fullerenes were first identified in the PN Tc~1 \cite{cami10}, and have been found in other Galactic and Magellanic Cloud PNe \cite{garcia-hernandez10, garcia-hernandez11, otsuka14}.  Because C-rich environments are necessary for fullerene formation, PNe exhibiting these features are also expected to be enriched in \emph{n}-capture elements.  Indeed, some of the largest \emph{s}-process enrichments have been found in fullerene PNe \cite{otsuka13a, sterling15, mashburn16, madonna18, aleman19}.  Intriguingly, there are exceptions to this trend, as IC~2501 \cite{sharpee07} and M~1-60 \cite{sterling15} do not appear to be \emph{s}-process enriched.  Many fullerene-bearing PNe lack sufficiently deep spectra for a detailed analysis of their heavy element abundances, and additional data are needed to better understand the evolutionary history of these objects.

\textbf{Fluorine -- another tracer of AGB nucleosynthesis.} Zhang \& Liu \cite{zhang_liu05} computed F abundances in a sample of PNe, and found that F and C abundances are correlated, in concurrence with studies of AGB stars \cite{jorissen92} and theoretical predictions.  AGB nucleosynthesis models indicate that C, F, and \emph{s}-process enrichments are largest for $\sim$2.5--3~M$_{\odot}$ stars \cite{cristallo15, karakas_lugaro16, karakas18}, while F is destroyed in more massive (M~$>5$~M$_{\odot}$) PN progenitors by proton captures during HBB \cite{mowlavi96}.  Otsuka \& Hyung \cite{otsuka_hyung20} strengthened the empirical evidence that \emph{s}-process and F enrichments in PNe are positively correlated (see also \cite{otsuka15, sterling15}), although more F detections are needed to quantify these correlations.  However, the subsolar F abundances in \emph{s}-process-rich PNe such as NGC~3918 \cite{zhang_liu05, garcia-rojas15} suggest that work is needed to improve the accuracy of nebular F abundances.

\textbf{Rb and the most massive PN progenitor stars.}  Garc\'ia-Hern\'andez et al.\ \cite{garcia-hernandez06, garcia-hernandez09} found large Rb enrichments in O-rich AGB stars with masses 4--8~M$_{\odot}$.  Although these abundances have been revised downward \cite{zamora14, perez-mesa17}, the enhanced Rb abundances combined with a lack of Zr enrichment indicate that \emph{s}-process neutrons are produced by $\alpha$-captures onto $^{22}$Ne in these stars.  The $^{22}$Ne source results in higher neutron densities than the $^{13}$C source, leading to branchings in the \emph{s}-process path that preferentially produce Rb \cite{vanraai12}.  The NIR [Rb~IV] \cite{sterling16} and [Rb~III] lines (Dinerstein et al.\ 2020a, in prep.) in PNe will enable more accurate Rb abundance determinations in PNe, and will help to identify which PNe are the descendants of the Rb-rich AGB stars studied by Garc\'ia-Hern\'andez and collaborators.

The discussion in this review has thus far neglected the role of binary PN progenitor stars, but this cannot be ignored since binary interactions likely play a role in the formation and shaping of a significant fraction of PNe \cite[e.g.,][]{demarco09, hillwig16, jones20b}, and can affect their chemical compositions \cite[e.g.,][]{jones14}.  For sufficiently close binaries, Roche lobe overflow can lead to common envelope (CE) evolution and the truncation of the AGB \cite{izzard06, karakas_lattanzio14}.  While the nebular compositions of only a few post-CE PNe have been studied, some show low C/O and N/O ratios that are consistent with a shortened AGB phase \cite{jones14}.  However, the final abundances depend on the evolutionary stage that the CE begins \cite{jones20a}.  Carbon and \emph{n}-capture element abundances in post-CE PNe can in principle be used to constrain which part of the AGB that CE onset occurred \cite{jones20a}, but it will be difficult to break the degeneracies with mass, metallicity, and other parameters affecting elemental yields \cite{karakas_lattanzio14}.

\section{Summary}

The advent of sensitive, high-resolution NIR spectrometers such as IGRINS, HPF, and iSHELL have played a critical role in the recent identification of several \emph{n}-capture element transitions.  These detections provide access to elements that cannot be observed at other wavelengths (e.g., Cd) and to ionization states that enable more accurate abundance determinations of Se, Br, Rb, Te, and Xe.  Studies of \emph{s}-process enrichments in extragalactic PNe are expected to grow, given the detections of \emph{n}-capture elements in Sagittarius Dwarf and Magellanic Cloud PNe, and the forthcoming launch of the \textit{James Webb Space Telescope}.  With the number of \emph{n}-capture elements that can be detected in PNe, it is possible to distinguish among AGB evolutionary models, which adopt different treatments of poorly-understood processes such as convection, mixing, and mass-loss, leading to predicted final abundances that can differ by factors of two or more \cite{karakas_lattanzio14, cristallo15, karakas_lugaro16}.  This provides crucial information for stellar yields at different metallicities, chemical evolution models, and constraining the production of \emph{n}-capture elements in all of their sites of origin.


\vspace{6pt} 




\funding{Results presented in this paper have received support from NSF awards AST~1715332 and AST~1412928.}

\acknowledgments{I thank the organizers of WorkPlaNS~II for their invitation, and K.\ F.\ Kaplan who reduced the HPF data shown in Figure~\ref{figure}.  This work includes preliminary results obtained with the Habitable-zone Planet Finder Spectrograph on the Hobby-Eberly Telescope, and the Infrared Telescope Facility (IRTF). The Hobby-Eberly Telescope is a joint project of the University of Texas at Austin, the Pennsylvania State University, Ludwig Maximilians-Universit\"at M\"unchen, and Georg-August Universit\"at Gottingen.  The IRTF is operated by the University of Hawaii under contract NNH14CK55B with the National Aeronautics and Space Administration.}

\conflictsofinterest{The author declares no conflict of interest.} 



\appendixtitles{no} 


\reftitle{References}


\externalbibliography{yes}
\bibliography{workplans.bib}





\end{document}